\documentclass[aps,prc,preprint,groupedaddress,showpacs,showkeys]{revtex4}
\usepackage{graphicx}
\usepackage{dcolumn}
\usepackage{bm}
\usepackage{epsfig}
\begin{document}
\bibliographystyle{apsrev}
%
%
%
\title{Higher Twist Effects in Parton Fragmentation Functions}
%
%
\author{Guanghua Xu,
Ed V. Hungerford, and
Larry Pinsky}
\affiliation{Department of Physics, University of Houston,
Houston, TX 77204}
%
%
%
\date{\today}
%
%
\begin{abstract}
We study twist expansions for parton fragmentation functions
based on the definition of the twist as an invariant matrix element of a
light-cone, bilocal operator.
The results are then applied to
a method which might be used to extract
higher twist effects
in the fragmentation sector using both
$e^+ e^-$, and $e^- p$ collisions.  We discuss how to apply the later
measurements to experiments at the Jefferson National Acceleration Facility.
\end{abstract}

%
\maketitle
%
%
%

\setcounter{equation}{0}
\section{Introduction}

Duality in the nucleon structure function, $ W_2 (\nu , Q^2 )$, was
discovered before Quantum Chromodynamics (QCD) \cite{blo}.
Thus the structure function, $ F_2 (Q^2 \omega^{\prime} ) = \nu W_2 / m_N$, 
in the resonance region ($W<2$GeV), is approximately the
same as (duals) the
deep inelastic region ($W>2$GeV), when the functions are
expressed in terms of
the scaling variable, $\omega^{\prime} = 1+W^2 /Q^2$.
Here $W$ is the final-state hadron mass.
Moreover, the
occurrence of duality appears to be local in the sense that it exists
in each interval of $\omega^{\prime}$ over the prominent nucleon
resonances.

An explanation of Bloom-Gilman duality was offered by de Rujula,
Georgi,
and Politzer in 1977 \cite{der}. Using the operator product expansion,
they represented the scaling function as a sum of various twist operators, and
then studied the contributions from individual twists  (moments of
the scaling functions) in the scaling variable,
$\xi = {2 x}/(1+(1+4x^2 m_N^2 /Q^2)^{1/2})$, where $x= Q^2 / 2 m_N \nu$.
They argue that the $n^{th}$ moment, $M_n (Q^2)$, of $F_2$ has the following
twist expansion,

\begin{eqnarray}
  M_n (Q^2)&  = & \sum^{\infty}_{k=1} ({m M_0^2 \over Q^2})^{k-1}
    B_{n, k} (Q^2 ) \, .
\end{eqnarray}

\noindent Here, $M_0^2$ is a mass scale $(\approx 400-500
MeV)^2$, and
$B_{n, k} (Q^2 )$ depends logarithmically on $Q^2$, being roughly on the
order of $B_{n, 0}$. According to eq. 1, there exists a region of $n$ and
$Q^2 (n \leq {Q^2 \over M_0^2 })$ where the higher twist contributions are
neither large nor negligible, and where the dominant contribution to the
moments comes from low-lying resonances. In this region for example, 
the moments defined by eq. 1 would not correspond to positive definite
functions. Thus one might expect the structure functions to oscillate
when large
$k$ is important, and this would lead to the appearance of local duality.
A more recent study of parton-hadron duality by Ji and Unrau \cite{jiu}
gives more quantitative estimations.

There is no doubt that duality is a very interesting phenomenon,
and it could allow one, under certain circumstances, to bridge
the gap between perturbative predictions and experimentally
observed quantities in non-perturbative regions \cite{pog}. However,
duality, as expressed in the above analyses, only reflects
properties of the parton structure functions. Appealing to the concept of
factorization in the strong interaction,
it is natural to ask whether there is duality in the parton fragmentation
sector. Although in twist-two, perturbative calculations, parton
structure functions
and parton fragmentation functions have some similarities, this
does not indicate that they have similar higher twist expansions.

Recently,
an experiment at the Jefferson National Acceleration Facility, Jlab
E00-108, studied duality in semi-inclusive
$ep$ reactions. Semi-inclusive reactions can reflect higher twist
contributions in the parton fragmentation sector, so it is worthwhile
to study these contributions in more detail in order to
understand how one might extract these contributions in such reactions.

We investigate these issues in this note. In Section 2, we
study the twist expansions in the fragmentation sector, and investigate
twist two perturbative calculations of semi-inclusive $ep$
reactions.  We also discuss a method to extract higher twist effects
in these measurements. The paper is summarized in Section 3.

\section{Twist Expansion in the Parton Fragmentation Sector}

In this section, we study the twist expansions of the parton
fragmentation functions in $e^+ e^-$ and $ep$ semi-inclusive reactions.

\subsection{$e^+ e^- \rightarrow h + X$}

\begin{figure}[htb!]
\begin{center}
\epsfxsize = 8cm
\epsffile{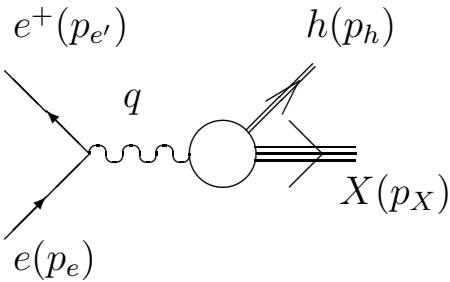}
\caption{Kinematics for $e^+ e^- \rightarrow h + X$.}
\label{fig:duality1}
\end{center}
\end{figure}

The kinematics for $e^+ e^- \rightarrow h + X$ is illustrated in
Fig. ~\ref{fig:duality1}.
Two kinematic invariants, $Q^2$ and $\nu = P \cdot q$, define the process.
We consider scalar particles, $h$.
The scattering matrix of the semi-inclusive reaction,
$e^+ e^- \rightarrow h+X$, when calculated to lowest order in the
electromagnetic interaction, is given by;

\begin{eqnarray}
 \langle hX|S_{QCD+em}|e^+ e^- \rangle
  & = & \langle hX|{\cal T}\, exp [i \int d^4 \xi ( {\cal
  L}_{em}^{int}
      +{\cal L}_{QCD}^{int} )] \, |e^+ e^- \rangle   \nonumber \\
    & = & \langle hX|{(ie)^2 \over 2!} \int d^4 \xi_1
          \int_{t_0}^{t} d^4 \xi_2
          {\cal T}[ {\cal L}_{em}^{int} (\xi_1) {\cal L}_{em}^{int}
  (\xi_2) ] \nonumber \\
     &  &     {\cal T} exp \, [i \int d^4 \xi {\cal L}_{QCD}^{int}] \,
          |e^+ e^- \rangle
          + {\cal O} (e^2)   \, ,
\end{eqnarray}

\noindent Here we have defined;

\begin{eqnarray}
  {\cal L} & = & e J^{\mu} A_\mu \, ,   \nonumber \\
  J^{\mu}_q (p_q ) & = & \sum_q Q_q : \bar{\psi}_q (p_{q'} )
    \gamma^{\mu} {\psi}_q (p_q ) : \, ,
\end{eqnarray}

\noindent where $Q_q$ is the charge of quark, $q$, in units of the
proton charge, $e$.

Using
Feynman rules for QED, one then finds for the kinematics of Fig.
~\ref{fig:duality1},
\begin{eqnarray}
  {\cal M} & = & {e^2 \over q^2} \bar{v}(p_{e'},\, \sigma_{e'}) \gamma_{\mu}
    u(p_e\, ,\, \sigma_e ) \langle hX|J^{\mu} (0)|0 \rangle \, .
\end{eqnarray}
\noindent Hence, the cross section can be written as;
\begin{eqnarray}
  d\sigma \sim \hat{l}_{\mu\nu} \hat{W}^{\mu\nu}
    {d^3 p_h \over (2\pi)^3 2E_h} \, ;
\end{eqnarray}
\noindent with;
\begin{eqnarray}
  \hat{l}_{\mu\nu} & = & {1\over 4} \sum_{\sigma_e\, \sigma_{e'}}
    \bar{v}(p_{e'},\, \sigma_{e'}) \gamma_{\mu} u(p_e\, ,\, \sigma_e )
    [\bar{v}(p_{e'},\, \sigma_{e'}) \gamma_{\nu} u(p_e\, ,\,\sigma_e )]^\dagger
    \nonumber \\
  &   & {1\over 2} \{ q_\mu q_\nu -q^2 g_{\mu\nu} -
          (p_e-p_{e'})_{\mu} (p_e-p_{e'})_{\nu} \} \, , \ \ \mbox{and}
\nonumber \\
  \hat{W}^{\mu\nu} & = & {1\over 4\pi} \sum_X (2\pi)^4
    \delta^4 (p_h + p_X -q) \langle 0|J^\mu (0) |hX \rangle
    \langle hX|J^\nu (0) |0 \rangle   \nonumber \\
  & = & {1\over 4\pi} \int d^4 \xi e^{iq \cdot \xi} \sum_X
    \langle 0|J^\mu (\xi) |hX \rangle \langle hX|J^\nu (0) |0 \rangle \, ,
\end{eqnarray}
\noindent We have set the electron mass to zero.
The sum over unobserved hadrons, $X$, cannot be complete because the state
$|hX \rangle$ depends non-trivially on the observed hadron. Therefore one does
not have $\sum_X |hX \rangle \langle hX| = 1$ and so
$\sum_X \langle 0|J_{\mu} (\xi) |hX \rangle \langle hX| J_{\nu} (0)|0 \rangle
\neq \langle 0|J_{\mu} (\xi)J_{\nu} (0)|0 \rangle$. Thus
$e^+ e^- \rightarrow h+X$ is not controlled by the
product of two operators, the operator product expansion does not apply,
and no short distance analysis can be formulated.

In a reference frame where the produced hadron, $h$, is fixed \cite{jef},
the hadron and photon momenta can be expanded as;

\begin{eqnarray}
  p_h^\mu & = & p^\mu + {m_h^2 \over 2} n^\mu   \nonumber \\
  q^\mu & = & {1\over m_h^2} (\nu -\sqrt{\nu^2 -m_h^2 Q^2}) p^\mu +
    {1\over 2} (\nu +\sqrt{\nu^2 -m_h^2 Q^2}) n^\nu \, ,
\end{eqnarray}
\noindent where;
\begin{eqnarray}
  Q^2 & \equiv & (p_e + p_{e'} )^2 \, ,   \nonumber \\
  p_h \cdot q & \equiv & \nu \, ,   \nonumber \\
  0<z & \equiv & {2p_h \cdot q \over q^2} \, ,   \nonumber \\
  p^\mu & = & {m_h \over 2} (1,\, 0,\, 0,\, 1)\, ,   \nonumber \\
  n^\mu & = & {1\over m_h} (1,\, 0,\, 0,\, -1) \, .
\end{eqnarray}
\noindent Writing;
\begin{eqnarray}
  \xi^\mu = \eta p^\mu + \lambda n^\mu + \xi^{\mu \perp} \, ,
\end{eqnarray}
\noindent one finds in the Bjorken limit;
\begin{eqnarray}
  \lim_{Q^2 \rightarrow \infty} q\cdot \xi = \eta \nu -{\lambda \over z} \, .
\end{eqnarray}
\noindent This result implies that as $\nu\rightarrow \infty$, then
$\eta\rightarrow 0$ and $\lambda \sim z$.
Because $\xi^2 = \xi_0^2-\xi_3^2 -\xi^{\perp\, 2} \leq \xi_0^2-\xi_3^2$,
one finds in the Bjorken limit that $\xi^{\mu \perp} \rightarrow 0$.
Therefore in this limit, light-like separation occurs, $\xi^\mu
\xi_\mu \sim 0$,
which dominates the integration region of eq. (6).

\begin{figure}[htb!]
\begin{center}
\epsfxsize = 8cm
\epsffile{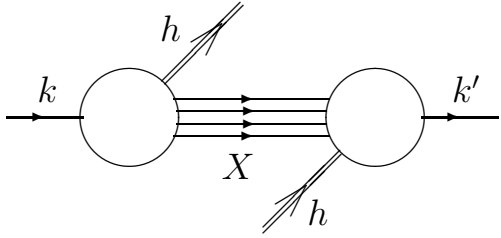}
\caption{Quark fragmentation.}
\label{fig:duality2}
\end{center}
\end{figure}

Fragmentation is generally a non-perturbative process. We first
consider the simplest quark fragmentation function represented
diagrammatically in Fig. ~\ref{fig:duality2}. More complicated
fragmentation processes,
such as coherent fragmentation of several quarks and gluons, do not
contribute until order $1/Q^2$.

\begin{figure}[htb!]
\begin{center}
\epsfxsize = 8cm
\epsffile{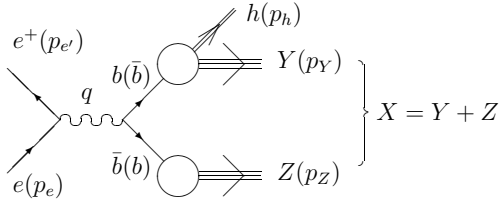}
\caption{Kinematics for $e^+ e^- \rightarrow h + X$ for
quark $b(\bar{b})$ with the Fig. ~\ref{fig:duality2} fragmentation.}
\label{fig:duality3}
\end{center}
\end{figure}

In the case of Fig. ~\ref{fig:duality2},
the diagram of $e^+ e^- \rightarrow h+X$ shown in
Fig. ~\ref{fig:duality1} can be re-drawn as in Fig. ~\ref{fig:duality3}.

The fermion field can be written as;
\begin{eqnarray}
  \psi (x) & = & \sum_s \int {d^3 p \over (2\pi)^{3/2}} {1 \over 2 E_p}
    [a(p,\, s) u(p,\, s) e^{-ip\cdot x} +
      b^{\dagger}(p,\, s) v(p,\, s) e^{ip\cdot x} ] \, ,
\end{eqnarray}
\noindent where $a(p,\, s)$ $(a^{\dagger}(p,\, s))$ and
$b(p,\, s)$ $(b^{\dagger}(p,\, s))$ are annihilation (creation) operators
for particles and anti-particles.
Assuming that there is no final state interaction between $Z$ and $hY$,
one can rewrite $\hat{W}^{\mu\nu}$ in eq. (6) as;
Fig. ~\ref{fig:duality2}  as:\\

\begin{eqnarray}
 \hat{W}^{\mu\nu} &  = & {1\over 4\pi} \int d^4 \xi e^{iq\cdot \xi}
    \sum_{Y,Z \{ Y+Z=X \} }
    \langle 0| J^\mu (\xi) |h(Y+Z) \rangle
    \langle h(Y+Z)| J^\nu (0) | 0 \rangle   \nonumber \\
    & = & {1\over 8\pi} \int d^4 \xi e^{iq\cdot \xi} \,
      \{  \nonumber \\
  &   & \sum_{Y,Z\{ Y+Z=X \}} : \langle 0| \bar{\psi}_\alpha (\xi) |hY \rangle
      \gamma^\mu_{\alpha\beta}
      \langle 0| \psi_\beta (\xi) |Z \rangle
      \langle Z|\bar{\psi}_{\delta} (0)|0 \rangle
              \gamma_{\delta\lambda}^{\nu}
              \langle hY| \psi_{\lambda} (0) |0 \rangle : +   \nonumber \\
    &   & \sum_{Y',Z'\{ Y'+Z'=X \} } :
      \langle 0| \bar{\psi}_{\alpha} (\xi) |Z' \rangle
      \gamma_{\alpha \beta}^{\mu} \langle 0| \psi_\beta (\xi) |hY' \rangle
      \langle hY'|
              \bar{\psi}_{\delta} (0)|0 \rangle
              \gamma_{\delta \lambda}^{\nu}
              \langle Z'|\psi_{\lambda} (0) |0 \rangle : \}   \nonumber \\
       & =  & {1\over 8\pi} \int d^4 \xi
              [ \, \sum_Y \langle 0| \bar{\psi}_\alpha (\xi) |hY \rangle
                \gamma_{\alpha \beta}^{\mu}
                \langle 0| \{ \psi_\beta (\xi),\, \bar{\psi}_\delta (0) \}
                  |0 \rangle \gamma_{\delta\lambda}^{\nu}
                \langle hY| \psi_{\lambda} (0)|0 \rangle +   \nonumber \\
       &   & \hspace{0.7in} \sum_{Y'}  \gamma_{\alpha \beta}^{\mu}
                \langle 0| \psi_\beta (\xi) |hY' \rangle
                \langle 0| \{ \bar{\psi}_\alpha (\xi),\, \psi_\lambda (0) \}
                  |0 \rangle
                \langle hY'| \bar{\psi}_\delta (0) |0 \rangle
                  \gamma_{\delta\lambda}^\nu \, ]   \, .
\end{eqnarray}
\noindent To obtain this result we have used the fact that;

\begin{eqnarray}
  \sum_Z |Z \rangle \langle Z| & = & 1 \ \ \mbox{and}  \nonumber \\
  \sum_{Z'} |Z' \rangle \langle Z'| &  = & 1
\end{eqnarray}
\noindent and due to the un-physical energy in the $|Z \rangle $ state,
then;
\begin{eqnarray}
   &  & \hspace{-0.2in} {1\over (2\pi)^4 }
    \int d^4 \xi e^{iq \cdot \xi} \hspace{-0.15in} \sum_{Z,Y\{ Y+Z=X \} }
    \hspace{-0.15in} \langle 0| \bar{\psi}_\alpha (\xi) |hY \rangle
    \gamma^\mu_{\alpha\beta} \langle 0| \bar{\psi}_\delta (0) |Z \rangle
    \langle Z| \psi_\beta (\xi) |0 \rangle \gamma^\nu_{\delta\lambda}
    \langle hY| \psi_\lambda (0) |0 \rangle   \nonumber \\
   & = &  \delta^4 ( p_Z+q -p_h -p_Y )
      \langle 0| \bar{\psi}_\alpha (0) |hY \rangle
    \gamma^\mu_{\alpha\beta} \langle 0| \bar{\psi}_\delta (0) |Z \rangle
    \langle Z| \psi_\beta (0) |0 \rangle \gamma^\nu_{\delta\lambda}
    \langle hY| \psi_\lambda (0) |0 \rangle   \nonumber \\
    &  =  &  0 \, ;  \nonumber \\
\noindent \mbox{and} \, ;  \nonumber \\
   &  & \hspace{-0.2in} {1\over (2\pi)^4 }
    \int d^4 \xi e^{iq \cdot \xi} \hspace{-0.15in} \sum_{Z,Y\{ Y+Z=X \} }
    \hspace{-0.15in}
    \gamma^\mu_{\alpha\beta} \langle 0| \psi_\beta (\xi) |hY \rangle
    \langle 0| \psi_\lambda (0) |Z \rangle
    \langle Z| \bar{\psi}_\alpha (\xi) |0 \rangle
    \langle hY| \bar{\psi}_\delta (0) |0 \rangle \gamma^\nu_{\delta\lambda}
    \nonumber \\
    & = & \delta^4 ( p_Z+q-p_h -p_Y )
    {\gamma^\mu}_{\alpha \beta} \langle 0| \psi_\beta (0) |hY \rangle
    \langle 0| \psi_\lambda (0) |Z \rangle
    \langle Z| \bar{\psi}_\alpha (0) |0 \rangle
    \langle hY| \bar{\psi}_\delta (0) |0 \rangle
    \gamma^\nu_{\delta\lambda} \nonumber\\
    & =  & 0 \, .
\end{eqnarray}
\noindent In this expression, summation over
color and flavor is assumed.

Due to the strong interaction Lagrangian, ${\cal L}_{QCD}^{int}$, eq. (12)
can be modified because of the higher order contributions in perturbative
QCD calculations. These interactions cause vertex corrections,
gluon polarization, etc, and
their contributions to the  cross section give logarithmic corrections.
These contributions do not change the
twists of the terms in the matrix elements. However, there are terms
which represent quarks propagating in a gluon background in order to
preserve color 
gauge invariance of the bilocal operator,
 $\hat{W}^{\mu\nu}$.  These terms are essential to generate higher
twist corrections, and can be included by changing the
singular function of the free field theory \cite{jef},
$\{ \psi (\xi),\, \bar{\psi} (0) \} = {1\over 2\pi} \not\partial \epsilon
(\xi_0) \delta (\xi^2 )$, to;
\begin{eqnarray}
  \{ \psi (\xi),\, \bar{\psi} (0) \} \rightarrow
    {1\over 2\pi} \not\partial \epsilon (\xi_0) \delta (\xi^2 )
    {\cal P} (exp [i \int_0^{\xi} d \zeta^{\mu} A_{\mu} (\zeta)] ) \, .
\end{eqnarray}

Substituting eq. (15) into eq. (12) and using the relationships;
\begin{eqnarray}
   \gamma^\mu \gamma^\rho \gamma^\nu & = &
    S^{\mu\rho\nu\sigma} \gamma_\sigma
    - i \epsilon^{\mu\rho\nu\sigma} \gamma_\sigma \gamma_5 \, ,
    \nonumber \\
  S_{\mu\rho\nu\sigma} \equiv {1\over 4}
    Tr(\gamma_\mu \gamma_\rho
    \gamma_\nu \gamma_\sigma ) & = &g_{\mu\rho} g_{\nu\sigma} +
    g_{\mu\sigma} g_{\nu\rho} - g_{\mu\nu} g_{\sigma\rho} \, ,
    \nonumber \\
  \langle 0| {\cal P} (exp \,[ i \int_0^{\xi} d \zeta^{\mu}
    A_{\mu} (\zeta) )] \, |0 \rangle
  & = & \sum_Z \langle 0| {\cal P}
    (exp \, [i \int_\infty^{\xi} d \zeta^{\mu} A_{\mu} (\zeta) )] \,
    |Z \rangle \cdot  \nonumber \\
    &   & \langle Z| {\cal P}
    (exp \, [i \int_0^\infty d \zeta^{\mu} A_{\mu} (\zeta) )] \, |0
    \rangle \, ,
\end{eqnarray}
\noindent one has,
\begin{eqnarray}
  \hat{W}^{\mu\nu} & = & {1\over 8\pi} \int d^4 \xi e^{iq\cdot \xi} \sum_X
    [ \, \langle 0| \bar{\psi}_\alpha (\xi) {\cal P}
    (exp\, [ - i \int_\xi^\infty d\zeta^\mu A_\mu (\zeta) )] \,  |hX \rangle
      {1\over 2\pi} \partial_\rho \epsilon (\xi_0) \delta (\xi^2 ) \cdot
    \nonumber \\
  &   & (S^{\mu\rho\nu\sigma} \gamma_\sigma -
      i \epsilon^{\mu\rho\nu\sigma} \gamma_\sigma \gamma_5 )_{\alpha\lambda}
    \langle hX| {\cal P} (exp \, [i \int_0^\infty d\zeta^\mu A_\mu
    (\zeta)] \,   )
      \psi_{\lambda} (0)|0 \rangle +   \nonumber \\
  &   & \langle 0| \psi_\beta (\xi) {\cal P}
    (exp \, [i \int_\xi^\infty d\zeta^\mu A_\mu (\zeta)] \, ) |hX \rangle
      {1\over 2\pi} [-\partial_\rho \epsilon ( \xi_0) \delta (\xi^2 )] \cdot
    \nonumber \\
  &   & (S^{\nu\rho\mu\sigma} \gamma_\sigma -
      i \epsilon^{\nu\rho\mu\sigma} \gamma_\sigma \gamma_5 )_{\delta\beta}
                \langle hX| {\cal P}
    (exp \, [ - i \int_0^\infty d\zeta^\mu A_\mu (\zeta)] \, )
      \bar{\psi}_\delta (0) |0 \rangle \, ]   \nonumber \\
  & = & {\cal P}_{h/\bar{b}} + {\cal P}_{h/b}   \, ,
\end{eqnarray}
\noindent where ${\cal P}_{h/b} ({\cal P}_{h/\bar{b}})$ is the fragmentation
function of a quark $b (\bar{b})$ fragmenting into hadron $h$.

As shown in eq. (10), light-like separation $\xi^{\mu} \xi_{\mu} \sim 0$
dominates the semi-inclusive process, $e^+ e^- \rightarrow h+X$ \cite{jef}.
Therefore one can write;
\begin{eqnarray}
  \xi^{\mu} & = & \lambda n^{\mu} + \hat{\xi}^{\mu} \, ,
\end{eqnarray}

\noindent where $\hat{\xi}^{\mu}$ includes contributions
from $\eta p^\mu$ and
$\xi^{\mu\perp}$. Both $\eta$ and $\xi^{\mu\perp}$ go to zero
($1/\sqrt{Q^2}$) as
$   Q^2 \to \infty   $

With eq. (8) and $n^2 = p^2 =0$, one can expand eq. (16) in terms of
$\hat{\xi}^{\mu }$. This gives,
\begin{eqnarray}
   \hat{W}^{\mu\nu} \hspace{-0.1in} &
    = & {1\over 4\pi} \int d \lambda \, e^{-i \lambda /z}
    \sum_X \, \{   \nonumber \\
    &  & \hspace{-0.1in} \sum_Y
    [ \, \langle 0| \bar{\psi}_\alpha (\lambda n) {\cal P}
    (exp \, [ - i \int_\lambda^\infty d\tau n\cdot A (\tau n)] \,  )
|hX \rangle
      {1\over 2\pi} \partial_\rho \epsilon (\xi_0) \delta (\xi^2 ) \cdot
    \nonumber \\
  &   & (S^{\mu\rho\nu\sigma} \gamma_\sigma -
      i \epsilon^{\mu\rho\nu\sigma} \gamma_\sigma \gamma_5 )_{\alpha\lambda}
    \langle hX| {\cal P} (exp \, [i \int_0^\infty d\tau n\cdot A (\tau
    n)] \,  )
      \psi_{\lambda} (0)|0 \rangle +   \nonumber \\
       &   & \langle 0| \psi_\beta (\lambda n) {\cal P}
    (exp \, [i \int_\lambda^\infty d\tau n\cdot A (\tau n)] \, ) |hX \rangle
      {1\over 2\pi} \partial_\rho \epsilon (-\xi_0) \delta (\xi^2 ) \cdot
    \nonumber \\
  &   & (S^{\nu\rho\mu\sigma} \gamma_\sigma -
      i \epsilon^{\nu\rho\mu\sigma} \gamma_\sigma \gamma_5 )_{\delta\beta}
                \langle hX| {\cal P}
    (exp \, [ - i \int_0^\infty d\tau n\cdot A (\tau n)] \,  )
      \bar{\psi}_\delta (0) |0 \rangle \, ]   \nonumber \\
   &   & + {\cal O}(\hat{\xi}       ) \} \, .
\end{eqnarray}

The $(\hat{\xi}      )^0$ term in eq. (19) is the same as the
definition of parton fragmentation given in ref.\cite{col}. Working in the
light-cone gauge, $n \cdot A = 0$, explicit reference to gluons
disappears, and
using the definition of the twist of an invariant matrix element of a
light-cone bilocal operator \cite{jef}, ref.\cite{xji}  studied the
twist expansion of the possible terms for the production of scalar hadron or
hadrons whose spins are not observable,
\begin{eqnarray}
  z\int {d\lambda \over 2\pi} e^{-i\lambda /z} \langle 0| \gamma^\mu
    \psi (0) |hX \rangle \langle hX| \bar{\psi} (\lambda n) |0 \rangle
  & = & 4[\hat{f}_1 (z) p^\mu + \hat{f}_4 (z) M^2 n^\mu ] \, ,  \nonumber \\
  &   &  \hspace{-3.95in}  \mbox{and}  \nonumber \\
  z\int {d\lambda \over 2\pi} e^{-i\lambda /z} \langle 0|
    \psi (0) |hX \rangle \langle hX| \bar{\psi} (\lambda n) |0 \rangle
  &  = & 4M \hat{e}_1 (z) \, .
\end{eqnarray}
\noindent Here $\hat{f}_1 (z) ,\, \hat{e}_1 ,\, \hat{f}_4$ have twists 2, 3, 4,
respectively; and $M$ is a generic QCD mass scale.

Since the twists for light-cone bilocal operators only represent the leading
$Q^2$ dependence, it is possible for
$\hat{f}_1 (z) ,\, \hat{e}_1$ and$\hat{f}_4$ to include multiplicative
factors of $M^2 /Q^2$.  Therefore eq. (20) can be
rewritten as;
\begin{eqnarray}
  &   & \hspace{-0.3in}
    z\int {d\lambda \over 2\pi} e^{-i\lambda /z} \langle 0| \gamma^\mu
    \psi (0) |hX \rangle \langle hX| \bar{\psi} (\lambda n) |0 \rangle
    = 4 \sum_{n=0}^\infty [\, p^\mu ({M \over \sqrt{Q^2}})^n \hat{f}_{1n} (z)
    + n^\mu ({M \over \sqrt{Q^2}})^{n+2} \hat{f}_{4n} (z) ];   \nonumber \\
  &   & \hspace{-0.3in} \mbox{and} \, ;  \nonumber \\
  &   & \hspace{-0.3in} z\int {d\lambda \over 2\pi} e^{-i\lambda /z} \langle 0|
    \psi (0) |hX \rangle \langle hX| \bar{\psi} (\lambda n) |0 \rangle
     =  4M \sum_{n=1}^\infty ({M \over \sqrt{Q^2}})^n \hat{e}_{1n} (z) \, .
\end{eqnarray}

One can see from eq. (17) that each $\hat{\xi} $ factor always has a
gauge-covariant derivative as a companion in the ${\cal O}(\hat{\xi}^\perp)$
terms; i.e. $\hat{\xi}^\mu D_\mu $. Using the light-cone,
$\hat{\xi}       \rightarrow {1/ \sqrt{Q^2}}$ as $Q \rightarrow \infty$ and
the twist analysis of bilocal operators \cite{jef}, one concludes
that the ${\cal O}(\hat{\xi}       )$ terms have higher twists than their
corresponding $(\hat{\xi}       )^0$ terms. For example, a general
${\cal O}(\hat{\xi}       )$ term of order $n$ has the following twist
expansion;
\begin{eqnarray}
  & & z\int {d\lambda \over 2\pi} e^{-i\lambda /z}
    \hat{\xi}_{\mu_1}       \cdot \cdot \cdot \hat{\xi}_{\mu_N}
    \Gamma^{\mu\nu}_{\alpha\beta}
    \langle 0| D^{\mu_1} \cdot \cdot \cdot D^{\mu_N}
    \psi_\alpha (\lambda n)|hY\rangle
  \langle hY| \bar{\psi}_\beta (0)|0 \rangle \nonumber\\
  & \sim & ({1\over \sqrt{Q^2}})^N \sum_{j = 0}^N
   p^{\mu_1} \cdot \cdot \cdot p^{\mu_{j}}
    n^{\mu_{j+1}} \cdot \cdot \cdot n^{\mu_N} M^{2(N-j)}
    [\, p^\mu p^\nu \hat{f}_{N+2}^{N-j} (z) +   \nonumber \\
  &   & (p^\mu n^\nu + n^\mu p^\nu ) M^2 \hat{f}_{N+3}^{N-j} (z) +
    n^\mu n^\nu M^4 \hat{f}_{N+4}^{N-j} (z) \, ]
    + \, trace\,\, terms  \, .
\end{eqnarray}
\noindent Based on the twist analysis from bilocal operators, the
leading twists
for $\hat{f}_{N+2}^{N-j}$, $\hat{f}_{N+3}^{N-j}$, and
$\hat{f}_{N+4}^{N-j}$ are $N+2+2(N-j)$, $N+3+2(N-j)$, and $N+4+2(N-j)$
respectively.

Since all the $trace\,\, terms$ have higher twists, their corresponding
diagonal terms, eq. (22), can be written as;
\begin{eqnarray}
  & & z\int {d\lambda \over 2\pi} e^{-i\lambda /z}
    \xi_{\mu_1}^\perp \cdot \cdot \cdot \xi_{\mu_N}^\perp
    \Gamma^{\mu\nu}_{\alpha\beta}
    \langle 0| D^{\mu_1} \cdot \cdot \cdot D^{\mu_N}
    \psi_\alpha (\lambda n)|hY\rangle
  \langle hY| \bar{\psi}_\beta (0)|0 \rangle
  \nonumber \\
  & \sim & \sum_{j = 0}^N p^{\mu_1} \cdot \cdot \cdot p^{\mu_{j}}
    n^{\mu_{j+1}} \cdot \cdot \cdot n^{\mu_N}
    [\, p^\mu p^\nu \sum_{i=0}^\infty
    ({M\over \sqrt{Q^2}})^{N+2(N-j)+i} \hat{f}_{N+2}^{N-j,\, i} (z) +
    \nonumber \\
  &   & \sum_{i=1}^\infty ({M\over \sqrt{Q^2}})^{N+2(N-j)+i}
    \hat{f}_{N+2}^{N-j,\, i} (z)
    + \sum_{i=2}^\infty ({M\over \sqrt{Q^2}})^{N+2(N-j)+i}
    \hat{f}_{N+2}^{N-j,\, i} (z) ] \, .
\end{eqnarray}

If there is final state interaction, there can be additional terms in
$\hat{W}^{\mu\nu}$. Ref. \cite{xji} has demonstrated
the existence of such a terms,
\begin{eqnarray}
  \hat{M}^\alpha_{\rho\sigma} & = & \int {d\lambda \over 2\pi}{d\mu \over 2\pi}
    e^{-i\lambda /z} e^{-i\mu (1/z_1 -1/z)} [\langle 0| i\vec{D}^\alpha_\perp
    (\mu n) \psi_\rho (0)|\pi (P) X \rangle \langle \pi (P) X |
    \bar{\psi}_\sigma (\lambda n) |0 \rangle   \nonumber \\
  &   & \hspace{-0.3in} +\int {d\lambda \over 2\pi}{d\mu \over 2\pi}
    e^{i\lambda /z} e^{i\mu (1/z_1 -1/z)} [\langle 0| \psi_\rho (\lambda n)
    | \pi (P) X \rangle \langle \pi (P) X| \bar{\psi}_\sigma (0)
    i{\stackrel{\leftarrow}{D}}^\alpha_\perp (\mu n) |0 \rangle \, .
\end{eqnarray}
In more complicated fragmentation processes,
more quarks and gluons are usually involved in the operators,
and these
lead to larger dimensions.
In any case, twist expansions can still be carried out based on a twist
analysis scheme involving bilocal operators.

In conclusion, one can carry out twist expansions for
all bilocal terms in eq. (19), to obtain the twist expansion for
$\hat{W}^{\mu\nu}$;
\begin{eqnarray}
  \hat{W} & = & \sum_n ({M^2 \over Q^2})^n f_n \, .
\end{eqnarray}

Twist two contributions to $|M|^2$ are the major contributions as
$Q^2 \rightarrow \infty$ or $({Q^2 \over M^2}) \gg n$.
As $({Q^2 \over M^2}) \geq n$, higher twist contributions become important.
Depending on the sign of the coefficients of $({Q^2 \over M^2})^n $,
the duality phenomenon in the fragmentation
sector could occur. Although one cannot carry out
non-perturbative calculations,
the determination of the sign of $\hat{f}_n$ could be measured in duality
experiments. Such experiments are straightforward in
$e^+ e^-$ collisions, where one can compare the cross sections of hadron, $h$,
production at fixed $z$ for various $Q^2$. For a given accuracy,
$e^+ e^-$ collisions at
lower $Q^2$ require higher-twist contributions than those at larger
$Q^2$.  However, it is meaningless
to attempt a twist expansion in the non-perturbative region, $n \geq
({Q^2 / M^2}) $.

Thus the fragmentation functions can also be expanded in terms of
contributions of operators of various twists, and the phenomenon of duality
can appear if the signs of the coefficients of the higher twist contributions
are not positive definite. Because complete
non-perturbative calculations are not presently possible, a comparison
of measurements of
fragmentation at various $Q^2$, having
the same $z$, can provide information
on the coefficients of higher twist contributions.

Although we have studied only the production of scalar hadrons, a
richer structure occurs if one
includes polarization effects. \cite{xji,jef}.

\subsection{The process $e^- p \rightarrow h+ X$}

\begin{figure}[htb!]
\begin{center}
\epsfxsize = 8cm
\epsffile{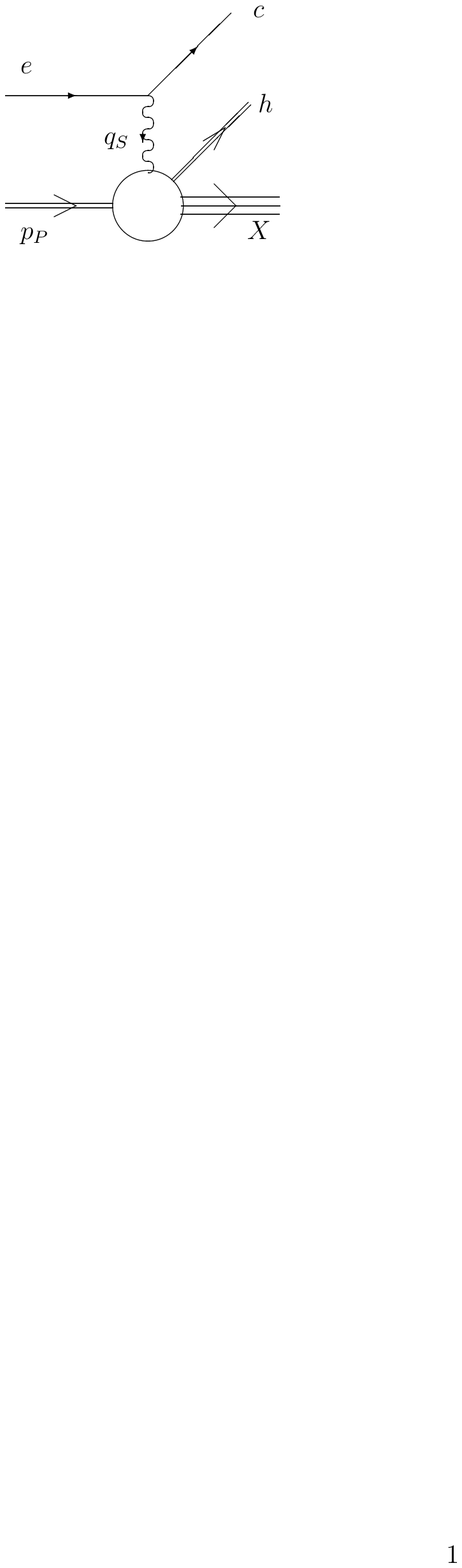}
\caption{Semi-inclusive $ep$ scattering.}
\label{fig:duality4}
\end{center}
\end{figure}

The scattering matrix of the process $e^- + p \rightarrow h+ X$ is given by,
\begin{eqnarray}
  W & = & \langle hX|T\, exp[i\int d^4 \xi\,
      ({\cal L}^{int}_{em}+{\cal L}^{int}_{QCD})] \, |ep \rangle \, .
\end{eqnarray}
The lowest order of the electromagnetic contribution is shown in
Fig. ~\ref{fig:duality4}.

\noindent The scattering matrix can be written as
\begin{eqnarray}
  W & = & \bar{\psi}_e (p^{\prime}_e ) \gamma^{\mu} \psi_e (p_e ) {1\over q^2}
      \langle hX| \hat{J}_{\mu} exp(i\int d^4 \xi {\cal L}^{int}_{QCD})
     |p \rangle + {\cal O} (e^2) \, ,
\end{eqnarray}
\noindent where $\hat{J}_{\mu}$ is the proton electromagnetic current,
and depends on the
proton structure.

From eq. (26), one has,
\begin{eqnarray}
  |W|^2 & = & \bar{\psi}_e (p^{\prime}_e ) \gamma^{\mu} \psi_e (p_e )
    \bar{\psi}_e (p^{\prime}_e ) \gamma^{\nu} \psi_e (p_e ) {1\over q^4}
    \sum_X \langle p| \hat{J}_{\mu}^{\dagger} |hX \rangle \cdot
    \langle hX| \hat{J}_{\nu} |p \rangle \, .
\end{eqnarray}

This is similar to
$e^+ e^- \rightarrow h+X$, where $\sum_X |hX \rangle \langle hX| \neq 1$
due to the
non-trivial dependence of $|hX \rangle$ on the observed hadron,
$h$. One can work out a
similar expansion as was done in the last subsection, to obtain the twist
expansion of eq. (28). The major difference between eq. (28) and eq. (5)
is that eq. (28) contains information on the parton structure functions
in addition to information on parton fragmentation functions.

The phenomenon
of duality in parton structure functions has been established through
inclusive $ep$ scattering. In order to extract higher twist information in
the parton fragmentation sector using semi-inclusive $ep$ scattering, one
needs to separate the contributions from the parton fragmentation 
and structure sectors.  If one allows factorization in the strong
interaction, one should be able to separate the scale, $Q_S^2$, 
in the structure
sector from the scale, $Q_F^2$, in the fragmentation sector.

In general, it is more advantageous to obtain higher twist information
in the
fragmentation sector using $e^+ e^-$ scattering. In order to extract
information from $ep$ scattering, one needs to subtract the contributions
from the structure sector. This can be achieved by suitably choosing
$Q_S^2$ and $Q_F^2$ in an experiment. One way to achieve this is to
keep the $Q_S^2$ sufficiently high so that the
contributions from the parton structure sectors are mainly from twist two
operators which can be confidently calculated. Then one can suitably choose
the $Q_F^2$ functions in the fragmentation sector to obtain information
of higher twist contributions. 
Therefore to quantitatively study higher-twist contributions in the parton
fragmentation sector, it is critical to identify $Q_S^2$
and $Q_F^2$ and separate the parton fragmentation sector from
the structure sector.

An experimental effort to detect higher twist effects in the fragmentation
sector through semi-inclusive $e^- p \rightarrow h+X$ scattering has
been recently undertaken at
Jlab. We comment on this process in more
detail here with the intent to clarify the experimental results.

The object of such an experiment would be to
determine higher twist effects by comparing measurements which
include effects from all twists to the calculated twist two results. 
As discussed previously, in order to study higher twist effects in
fragmentation sectors through
semi-inclusive $e^- p \rightarrow h+X$ scattering, it is important to
subtract the contributions from the parton structure functions.
This is feasible by choosing high $Q_S^2$ in the
structure sector so that these contributions are predominantly
from twist two operators which can be calculated perturbatively by using the
well-known information on twist two parton structure functions. One must
have measurements in a sufficiently high
region of $Q_S^2$, with various $Q_F^2$, in order to compare the
measurements with calculations of twist two contributions in the
same regions. The differences between the measurements and the calculations
provide information of higher twist effects in the fragmentation
sector. Therefore we need to locate the regions with high $Q_S^2$ and
various $Q_F^2$ to carry out the twist-two calculations.

\begin{figure}[htb!]
\begin{center}
\epsfxsize = 8cm
\epsffile{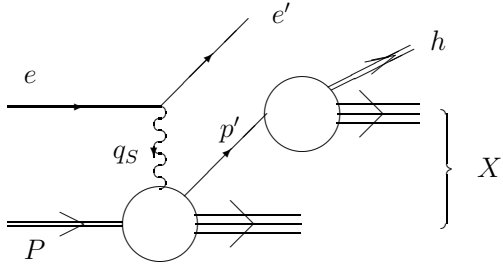}
\caption{Semi-inclusive $ep$ scattering Kinematics for $e^+ e^-
  \rightarrow h + X$.}
\label{fig:duality5}
\end{center}
\end{figure}

If $Q_S^2$ is large enough so that the contributions from the parton
structure sectors are twist two, the scattering process
$e^- p \rightarrow h+X$ is illustrated in Fig. ~\ref{fig:duality5}.

\noindent Before presenting a calculation of the process shown in
Fig. ~\ref{fig:duality5}, we would like to make several remarks.
\begin{enumerate}
\item In a momentum infinitive reference frame where
$|\vec{p}| \gg m_p$, parton momentum
and energy are usually expressed as a fraction $x$ of the momentum
(energy) of
the parent proton momentum (energy), where $x={Q^2 \over 2p\cdot p_e}$. In the
CMS or Lab frame, a Jlab
beam energy ($E_e \sim 5.5\, GeV$) is too low
to use $x={Q^2 \over 2p\cdot p_e}$ as the momentum fraction
since the beam momentum
$p_e$ in both CMS and Lab does not satisfy $|\vec{p}_e| \gg m_p$. In our
approach in this section, we work in the target rest frame. While the target
mass, $m_p$, is non-negligible when compared to the beam energy, the
quark masses
are. Therefore, the quark energy and momentum in the target rest frame can be
expressed as;
\begin{eqnarray}
  E_q & = & x m_p \, ,   \nonumber \\
  | \vec{p}_q | & = & x m_p \, .
\end{eqnarray}

In a finite momentum reference frame, parton transverse momentum
cannot be ignored and the parton structure function $f_q$ should depend on $x$,
$Q^2$, and $p_q^T$, i.e. $f_q = f_q (x,Q^2,p_q^T)$, where $x = E_q/E_{targ}$.
This structure function should be the same as the one measured in an infinite
momentum reference frame, if one
boosts the reference frame to a very high momentum so that $p_q^T$ is
negligible. In a target-rest reference frame, the probability for
a parton to have a fraction $x$ of the parent proton's energy should be the
same as the corresponding parton structure function measured in
momentum-infinitive reference frame because $x = E_q/E_{targ}$. However,
the direction of the momentum is completely random and satisfies
$\sum_q \vec{p}_q \equiv 0$.
\item In $e^+ e^-$ scattering, the $Q_F^2$ is defined as
$Q_F^2 = (p_{e^+} +p_{e^-})^2$ and is equal to $4E_{beam}^2$ in the CMS.
If one concentrates on the fragmentation functions given by
Fig. ~\ref{fig:duality2},  the
energy of the produced quarks, $E_q$, is equal to $E_{beam}$ in the
CMS. Therefore
$Q_F^2$ can be written as;
\begin{eqnarray}
  Q_F^2 & = & 4E_q^2 \, .
\end{eqnarray}
The fraction of energy carried by the produced hadron from the parent
parton is $z=2p_h \cdot q / q^2$, which is equal to $E_h /E_q$ in the CMS.

In semi-inclusive $e^- p$ scattering, one usually defines (see
Fig. ~\ref{fig:duality4})
\cite{clo}
\begin{eqnarray}
  Q_F^2 & \equiv & -q_S^2 \, ,   \nonumber \\
  z & \equiv & {p_P \cdot p_h \over p_P \cdot q}  \nonumber \\
  & = & \left( {E_h \over q^0} \right)_{lab.\,\,frame} \, .
\end{eqnarray}
In order to adopt the fragmentation functions obtained from $e^+ e^-$
scattering in the $e^- p$ scattering process, i.e.
\begin{eqnarray}
  {\cal P}^{ep}_{h/q} (Q_F^2, z) & = & {\cal P}^{ep}_{h/q} (Q_F^{\prime 2},
   z')  \nonumber \\
   & = &{\cal P}^{e^+ e^-}_{h/q} (Q_F^{\prime 2}, z^\prime )  \, ,  \nonumber
\end{eqnarray}
one should use the same
definitions used in $e^+ e^-$ scattering. Therefore,
we correspondingly redefine the $Q_F^2$ and $z$ in $e^- p$ scattering as
\begin{eqnarray}
  z & \equiv & {p_P \cdot p_h \over {p_P} \cdot p_{q'}}  \nonumber \\
    & = & \left( {E_h \over E_{q'}} \right)_{lab.\,\,frame} \, ,   \nonumber \\
  Q_F^2 & \equiv & {4( p_P \cdot p_{q'})^2 \over p_P^2 }  \nonumber \\
    & = & \left( 4E_{q'}^2 \right)_{lab.\,\,frame} \, .
\end{eqnarray}
Compared to $Q_F^2$ and $z$ in $e^- e^+$ scattering, we believe that
these definitions are reasonable and the new $Q_F^2$ is at least proportional
to $Q_F^2$ in $e^- e^+$.
\item The $Q_S^2$ in $ep$ scattering is defined as usual as;
\begin{eqnarray}
  Q_S^2 = -q_S^2 & = & -(p_e-p_{e'})^2 \simeq 2E_e E_{e'} (1-\cos
  \theta_{ee'}) \, .
\end{eqnarray}
\end{enumerate}

The differential cross section of the process $e^- p\rightarrow h+X$ is
given by,
\begin{eqnarray}
  d\sigma (e^- p\rightarrow h+X) & = & \sum_q \int dx f_q (x)
    {d\Omega_q \over 4\pi } {1 \over |\vec{v}_e -\vec{v}_q | 2E_e 2E_q}
    |{\cal M}|^2 {1 \over (2\pi)^2} \cdot  \nonumber \\
  &   & \delta^4 (p_{e'}+p_{q'}-p_e-p_q)
    {d^3 p_{e'} \over 2E_{e'}}{d^3 p_{q'} \over 2E_{q'}} D_q^h (Q_F^2, z)
    dz \, ,
\end{eqnarray}
where $p_e$, $p_{e'}$, $p_q$, and $p_{q'}$ are four momenta of the incoming
electron, scattered electron, quark within a target proton, and scattered
quark, respectively.  Also $f_q (x)$ is the structure function of quark $q$ in
a target proton with $x$ fraction of the proton energy, $D_q^h (Q_F^2, z)$
is the fragmentation function of quark $q$ fragmenting to hadron $h$ with
$z$ fraction of the quark $q$ energy, $\Omega_q$ is the solid angle of
quark $q$ within the target in the lab frame, and the $\delta$-function
reflects energy-momentum conservations among $p_e$, $p_{e'}$, $p_q$, and
$p_{q'}$.

We define the electron beam direction as the positive $z$ direction and let;
\begin{eqnarray}
  W_1 & = & E_e E_{q'} (1- \cos \theta_{q'}) +
    {Q^2_S \over 2E_e} (E_e - E_{q'} \cos \phi_{q'} ) \, ,   \nonumber \\
  W_2 & = & W_1 E_{q'} (1-\cos \theta_{q'} ) +
    {Q^2_S \over 2E_e} E_{q'}^2 \sin^2 \theta_{q'}
    \cos^2 (\phi_{e'}-\phi_{q'}) \, ,   \nonumber \\
  W_3 & = & W_1^2 + ({Q^2_S \over 2E_e})^2 E_{q'}^2 \sin^2 \theta_{q'}
    \cos^2 (\phi_{e'}-\phi_{q'}) \, .
\end{eqnarray}
Eq. (34) can be rewritten as;
\begin{eqnarray}
  {d\sigma (e^- p\rightarrow h+X) \over dQ_S^2 \, dQ_F^2 \, dz} & = &
    {1 \over (16 \pi)^3 m_P^2} \sum_q
    \int d\phi_{e'} d\cos\theta_{q'} d\phi_{q'}
    {1 \over |\vec{v}_e -\vec{v}_q | E_e^2 E_q} f_q (x) \cdot  \nonumber \\
  &   & {1 \over x^2 (1-\cos\theta_{qe'})}
    |{\cal M}|^2 D_q^h (Q_F^2, z) \, ,
\end{eqnarray}
where
\begin{eqnarray}
  E_{e^\prime} & = & \left\{ \begin{array}{r@{\quad\quad}l}
   \frac{\displaystyle W_2 + \sqrt{W_2^2 -W_3 E_{q'}^2 (1-\cos\theta_{q'})^2}}
    {\displaystyle E_{q'}^2 (1-\cos\theta_{q'})^2} &
    if \ \cos(\phi_{e'} -\phi_{q'}) \geq 0   \\
    \\
   \frac{\displaystyle W_2 - \sqrt{W_2^2 -W_3 E_{q'}^2 (1-\cos\theta_{q'})^2}}
    {\displaystyle E_{q'}^2 (1-\cos\theta_{q'})^2} &
    if \ \cos(\phi_{e'} -\phi_{q'}) < 0  \end{array} \right.  \, ,
    \nonumber \\
  x & = & {E_{e'} +E_{q'} -E_e \over m_P} \, ,   \nonumber \\
  \cos \theta_q & = & {E_{e'} \cos \theta_{e'} + E_{q'} \cos \theta_{q'} -E_e
\over x m_P } \, ,   \nonumber \\
  \sin \theta_q & = & {\sqrt{E^2_{e'} \sin^2 \theta_{e'} + E^2_{q'} \sin^2
\theta_{q'} + 2E_{e'} E_{q'} \sin \theta_{e'} \sin \theta_{q'} \cos(\phi_{e'}
-\phi_{q'}) } \over x^2 m^2_P} \, ,   \nonumber \\
  \cos \phi_q & = & {E_{e'} \sin\theta_{e'} \cos\phi_{e'} + E_{q'}
\sin\theta_{q'} \cos\phi_{q'} \over xm_P \sin\theta_q} \, ,   \nonumber \\
  \sin \phi_q & = & {E_{e'} \sin\theta_{e'} \sin\phi_{e'} + E_{q'}
\sin\theta_{q'} \sin\phi_{q'} \over xm_P \sin\theta_q} \, .
\end{eqnarray}

\begin{figure}[htb!]
\begin{center}
\epsfxsize = 8cm
\epsffile{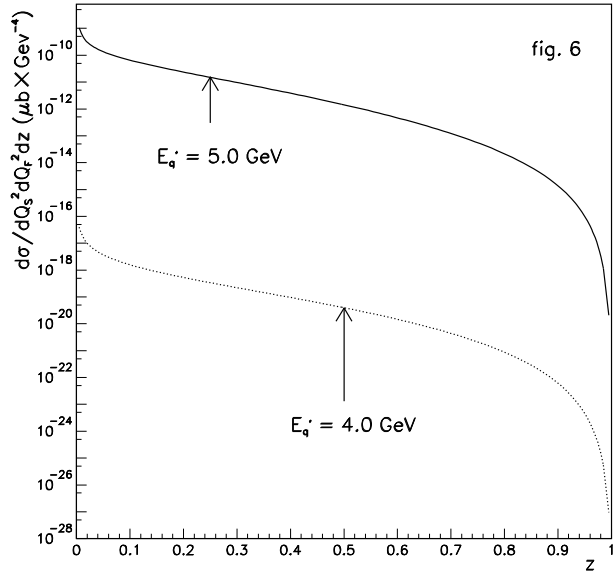}
\caption{Calculations of $d\sigma / dQ_S^2 dQ_F^2 dz$ versus $z$ for
$Q_S^2 = 10.0\, GeV^2$ and $E_{q'} = 5.0\, GeV$ and $4.0\, GeV$.}
\label{fig:duality6}
\end{center}
\end{figure}

Using the parton structure functions from ref. \cite{cb} and the pion
fragmentation functions from ref. \cite{pc}, the pion production cross
sections versus z at $Q_S^2 = 10.0\, GeV^2$ and $E_{q'} = 5.0\, GeV$ and
$4.0\, GeV$ are plotted in Fig. ~\ref{fig:duality6}. In general, the
production cross
section is small and the choices of $E_{q'}$ that satisfy all conditions
are limited. This can be seen from the very small cross section at
$E_{q'} = 4.0\, GeV$ in Fig. ~\ref{fig:duality6} when $Q_S^2 = 10.0\, GeV^2$.
This problem can be avoided by lowering the beam
energy, which leads to lower $Q_S^2$ and more choices of $E_{q'}$.
However, lower $Q_S^2$ brings more
contributions from higher twist operators in the structure sector. 
When
the magnitude of $Q_S^2$ is sufficient that the parton structure
functions are safely represented by twist 2 operators, one can compare
the hadron production data at various $E_{q'}$ to obtain information
on higher twist contributions in the fragmentation sector. 

The expressions given in eq. (34) are given  for the purpose of calculating the
hadron production cross section versus $z$ at $Q_S^2$ and $E_{q'}$.
In experiments, the quantities that can be directly measured are $E_{e'}$,
$\theta_{e'}$, $\phi_{e'}$, $E_h$, and approximately $\theta_{q'}$ and
$\phi_{q'}$, if one takes $\theta_{q'} \approx \theta_h$ and
$\phi_{q'} \approx \phi_h$. The
quantities that can not be directly measured are $x$, $\theta_q$, $\phi_q$,
$E_{q'}$, and $z$. One can determine $x$, $\theta_q$, $\phi_q$, and $E_{q'}$
based on energy and momentum conservation and the measurements of $E_{e'}$,
$\theta_{e'}$, $\phi_{e'}$, $\theta_{q'}$, and $\phi_{q'}$ with the
following equations,
\begin{eqnarray}
  Q^2_S & = & 2E_e E_{e'} (1-\cos\theta_{e'}) \, , \nonumber \\
  E_{q'} & = & {E_e E_{e'} (1-\cos\theta_{e'}) \over
    E_{e'} (1-\cos\theta_{e'q'}) - E_e (1-\cos\theta_{q'}) } \, , \nonumber \\
  x & = & {E_{e'} + E_{q'} - E_e \over m_P } \, , \nonumber \\
  \cos\theta_e & = & {E_{e'} \cos\theta_{e'} + E_{q'} \cos\theta_{q'} - E_e
    \over m_P } \, , \nonumber \\
  \sin\theta_e & = & {(E_{e'}^2 \sin^2 \theta_{e'} +
    E_{q'}^2 \sin^2 \theta_{q'} +
    2E_{e'} E_{q'} \sin\theta_{e'} \sin\theta_{q'} \cos (\phi_{e'} -\phi_{q'}
    )^{1/2} \over m_P } \, , \nonumber \\
  \cos\phi_e & = & {E_{e'} \sin \theta_{e'} \cos\phi_{e'}
    + E_{q'} \sin \theta_{q'} \cos\phi_{q'}
    \over m_P \sin\theta_e \cos\phi_e } \, , \nonumber \\
  \sin\phi_e & = & {E_{e'} \sin \theta_{e'} \sin\phi_{e'}
    + E_{q'} \sin \theta_{q'} \sin\phi_{q'}
    \over m_P \sin\theta_e \sin\phi_e } \, , \nonumber \\
  \cos\theta_{e'q'} & = & \cos\theta_{e'} \cos\theta_{q'} +
    \sin\theta_{e'} \sin\theta_{q'} \cos (\phi_{e'} -\phi_{q'} ) \, .
\end{eqnarray}
Also $z$ can be determined through the above determination of $E_{q'}$ and
the measurements of $E_h$,
\begin{eqnarray}
  z & = & { E_h \over E_{q'} } \, .
\end{eqnarray}

With eqs. (38) and (39), one can determine $Q^2_S$, $Q^2_F$, and $z$ of the
hadron production processes and then obtain the cross section versus $z$.
Comparing the cross sections at various $Q^2_F$ but with the same $Q^2_S$
and $z$, one can deduce the higher twist contributions from the fragmentation
sector.

\section{Summary}

We have studied possible  higher twist QCD contributions in the
fragmentation sector. In summary we find the following.

\begin{enumerate}
\item The contributions to $e^- e^+ \rightarrow h+X$ cross section can be
expanded as contributions of operators with various twists. If
$Q_F^2 \rightarrow \infty$, twist-two operators are the dominant
contribution. As one decrease $Q_F^2$, higher-twist contributions become
more and more important. If the sign of the coefficients in higher-twist
contributions are
not positive definite, oscillation similar to duality in $e^- p$ inclusive
scattering could appear. As $Q_F^2$ becomes low, QCD becomes completely
non-perturbative, and the twist expansion becomes meaningless.

\item A study of higher-twist contributions can be carried out through
$e^+ e^- \rightarrow h+X$  by comparing measurements of
cross sections versus $z$ at various $Q_F^2$. In order to carry out the
same study using $e^- p \rightarrow h+X$, one needs to make sure that
the higher twist contributions from the parton structure sector are
well determined, so that one can isolate the contributions from the
fragmentation sector. One way to achieve this is to choose large enough
$Q_S^2$ so that the contributions to the cross section
from the structure sectors are dominantly from twist-two operators,
which can be calculated with well measured structure functions and
perturbation QCD.
To achieve higher $Q_S^2$ but different $Q_F^2$,
one can suitably choose the detector angles carefully in
$e^- p \rightarrow h+X$ experiments for various beam energies.
\item We have not attempted to study the evolution of the
ratio,$\sigma_{e^- p \rightarrow hX} (Q_S^2) /\sigma_{tot} (Q_S^2)$ as
a function of $z$.  However we suggest that this ratio contains
information of the higher twist effects in the fragmentation sector
and should be studied in the future.
\item To adopt the fragmentation functions obtained in
$e^- e^+ \rightarrow h+X$ in the process of $e^- p \rightarrow h+X$,
redefinitions of $Q_F^2$ and $z$ are necessary.
We currently use the definitions in eq. (31).
\end{enumerate}

More studies on this subject, especially considering the case of polarization,
are necessary.

\noindent{\Large\bf Acknowledgements}

This research was supported in part by the US Department of Energy under
the grant DE-AS05-76ERO-3948.

\end{document}